# Disk Wind Mass Loss Estimates in QSOs

Nicolas A. Pereyra[*]
pereyrana@utpa.edu
Department of Physics and Geology, University of Texas – Pan American

**Abstract**

We derive here a relatively simple expression for the total wind mass loss rates in QSOs within the accretion disk wind scenario. We show that the simple expression derived here for QSO disk wind mass loss rate is in very good agreement with the more "exact" values obtained through significantly more complex and detailed numerically intensive 2.5D time-dependent simulations. Additionally we show that for typical QSO parameters, the disk itself will be emitting mostly in the UV/Optical spectrum, in turn implying that the X-ray emission from QSOs likely is produced through some physical mechanism acting at radii smaller than the inner disk radius (for a standard accretion disk, half of the initially gravitational potential energy of the accreting disk mass is emitted directly by the disk, while the other half "falls" closer towards the black hole than the inner disk radius). We also show that for typical QSO parameters, the disk itself is dominated by continuum radiation pressure (rather than thermal pressure), resulting in a "flat disk" (except for the innermost disk regions).

[*] Corresponding author: Fax 1 956 665 2423
E-mail address: pereyrana@utpa.edu (Nicolas A. Pereyra)

# Estimadas de Perdidas de Masa por Vientos en Quásares

Nicolas A. Pereyra[*]
pereyrana@utpa.edu
Department of Physics and Geology, University of Texas – Pan American

**Abstract**

Derivamos aquí una expresión relativamente simple para el cálculo de la pérdida total de masa por viento de disco en quásares dentro de el escenario de vientos de discos de acreción. Mostramos que la expresión simple que derivamos aquí para la pérdida de masa en discos de vientos en quásares esta en buen acuerdo con los resultados mas "exactos" obtenidos a través de modelos significativamente más complejos y detallados de simulaciones hidrodinámicas de 2.5 dimensiones numéricamente intensivas. Adicionalmente, mostramos que para parámetros típicos de quásares, el disco mismo está emitiendo casi complemente en la parte ultravioleta/óptica del espectro, as su vez implicando que la emisión de rayos-x en quásares es emitido a través de algún mecanismo físico en una región de radio menor que el radio interno del disco (para un disco estándar, la mitad de la energía potencial gravitatoria de la masa del disco es emitida directamente por el disco, mientras que la otra mitad de la energía "cae" en una región dentro del radio interno del disco). Nosotros aquí también mostramos que para parámetros típicos de quásares, el disco mismo es dominado por presión de radiación continua (en vez de presión termal) resultando en un disco "plano" (excepto por la región más interna del disco).

*Palabras Claves:* Acreción, Discos de Acreción, Hidrodinámica, Quásares

---

[*] Autor de contacto: Fax 1 956 665 2423
Correo Electrónico: pereyrana@utpa.edu  (Nicolas A. Pereyra)

## 1. Introduction

Positions of "Quasi-Stellar Objects" or QSOs (also referred to as quasars) were initially reported by the identification of the optical counterparts of strong radio sources (e.g. Griffin [18]; Read [41]; Sandage et al. [44]) These objects were observed to have the following characteristics: large ultravioled radiation flux, broad emission lines, and spectrum lines with high redshifts (e.g., Greenstein et al. [17]; Mathews et al. [31]; Greenstein et al. [16]; Ryle et al. [42]; Schmidt et al. [47]; Burbidge [7]; Sandage [43]; Schmidt [46]; Bahcall et al [4]; Oke [35]). Also a systematic search for QSOs based upon the excess of ultraviolet radiation resulted in the detection of radio-quiet QSOS (Sandage [43]).

Some QSOs were also observed to present broad absorption lines (BAL) within P-Cygni type line profile (non-shifted emission line superimposed with a blue-shifted absorption line) in the UV rest frame (e.g., Lynds [29]; Burbidge [8]; Burbidge [9]; Scargle et al. [45]) Comparing the absorption troughs of theoretical P-Cygni profiles derived from resonance scattering from a spherical wind with the P-Cygni profiles observed in BAL QSOs, Lucy [28] suggested that the flows generating the BALs were not spherically symmetric. Mushotzky et al. [33] suggested that the BAL QSOs could be caused by intrinsic flows of the QSOs driven by line radiation pressure.

Bowyer et al [6], with well collimated proportional counter flown on an Aerobee 150 Rocket, detected X-ray emission from a QSO. During the decade of the 1970s NASA launches several X-ray Observatories: UHURU, HEAO-1, and Einstein Observatory; observations from these X-ray telescopes showed that X-ray emission is a common property of QSOs, although not all QSOs are observed to be strong X-ray emitters (e.g., Zamorini et al [54]).

Further observations with ground telescopes led to the detection of BAL QSOs with detached absorption troughs (Osmer et al. [36]) and BAL QOSs with multiple absorption troughs (Turnshek et al. [51]). Turnshek et al. [51] suggested that most QSOs are ejecting high velocity flows, but that the covering factor was relatively low so that only a fraction of the observation angles would detect BALs. Calculating theoretical line profiles due to resonance scattering for models with various covering factors, Junkkarinen [23] found that the form of the absorption troughs in P-Cygni BALs could be accounted for in systems that presented absorption regions with small covering factors. Drew et al [12] found that X-ray luminosities comparable to UV luminosities would ionize the wind to a point where the populations of ions responsible for the BALs observed in QSO would be too low to produce observable absorption lines; Drew et al. [12] suggested that BAL QSOs are either intrinsically X-ray quiet or the flows generating the BALs are X-ray shielded.

Junkkarinen et al. [24] through the analysis of the spectrum of several BAL and non-BAL QSOs, found that the emission line and continuum properties of non-BAL and BAL QSOs were closely similar; consistent with the view that BAL and non-BAL QSOs are intrinsically the same kind of objects whose apparent differences are due to varying observation angles. Arav et al. [2] and Arav et al [3] studied line radiation pressure as a mechanism to accelerate flows in QSOs and they found that line-radiation pressure could account for the terminal velocities of the order of $10^4$ km s$^{-1}$, consistent with the velocities inferred in from BAL QSOs. Murray et al. [32] developed 1D dynamical models of accretion disks from QSOs. More recently, Pereyra and collaborators (Hillier et al. [22], Pereyra [38]) developed 2.5D time-dependent hydrodynamic simulations of QSO accretion disk winds.

Murray et al. [32] showed through 1D dynamical models that, with an X-ray shielding mechanism, and accretion disk wind driven line-radiation pressure with streamlines approximately parallel to the disk at high velocities could roughly account for: the flow velocity inferred in BAL QSOs, the detached troughs observed in many QSOs, and for the percentage of observed BAL QSOs with respect to QSOs in general > 10% (Hewett et al. [21]), interpreting the detections of broad absorption lines in QSOs as dependent on viewing angle.

Pereyra and collaborators (Hillier et al. [22], Pereyra [38]), through more realistic 2.5D time-dependent hydrodynamic simulations of QSO accretion disk winds confirmed the results of Murray et al. [32]. Further Pereyra and collaborators, not only confirmed that the accretion disk wind (ADW) scenario could account for the percentage of BALs of QSOs and for the BAL QSOs with detached troughs, but they showed that the ADW scenario could also account for the multiple absorption troughs observed in many BALS QSOs as a consequence of viewing angle; specifically, multiple absorption troughs could be observed at viewing angles that were close to the angle that separates the X-ray shielded region and the X-ray non-shielded region. Pereyra and collaborators found that in the ADW scenario multiple absorption troughs would be a consequence of the discontinuity of wind ionization balances (rather than discontinuities in the density/velocity structure of the outflow as has been suggested previously by many different authors).

Thus the ADW scenario for QSOs, suggested early on by Turnshek [51], studied in detailed first by Murray et al. [32] through 1D dynamical models, and more recently studied with more realistic 2.5D time-dependent hydrodynamic simulations by Pereyra and collaborators (Hillier et al. [22], Pereyra [38]), continues to be a promising scenario to account for BAL QSOs as a consequence of viewing within a unified model for QSOs.

In this work we derive a relatively simple expression to estimate wind mass loss rates from QSOs within the ADW scenario and compare it with the significantly more complex approach of developing a 2.5D time-dependent simulations and numerically integrating over the disk surface once a quasi-steady solution is found (Hillier et al. [22], Pereyra [38]). As we show below, we find very good agreement with results obtained through the approximate expression derived here, and the more "exact" results obtained through the detailed numerically intensive 2.5D time-dependent models of Pereyra and collaborators.

**2. Standard Accretion Disk Model**

In the standard accretion disk model (Shakura et al. [48]), the disk is assumed to be in a steady state. Shear stresses transport angular momentum outwards as the material of the gas spirals inwards. Conservation of angular momentum leads to the following expression

$$W 2\pi r^2 - \dot{M}_{accr} \omega r^2 = C \text{ (constant)} \qquad (1)$$

where $r$ is the radius, $\dot{M}_{accr}$ is the mass accretion rate, $\omega$ is the corresponding angular velocity and $W$ is defined by

$$W \equiv \int_{-z_0}^{z_0} w_{r\phi} dz \qquad (2)$$

where $z_0$ is the half thickness of the disk and $w_{r\phi}$ is the shear stress between adjacent layers. With the additional assumption that the shear stresses are negligible in the inner disk radius

$$W = \frac{\dot{M}_{accr}}{2\pi r^2}(\omega r^2 - \omega_i r_i^2) \qquad (3)$$

where $\omega_i$ is the angular velocity at the inner disk radius $r_i$. Taking into account the work done by shear stresses, and assuming that, as the mass accretes inwards, the gravitational energy lost is emitted locally

$$Q = \frac{1}{4\pi r}\frac{d}{dr}\left[\dot{M}_{accr}\left(\frac{\omega^2 r^2}{2} - \frac{GM_{bh}}{r}\right) - W2\pi\varpi r^2\right] \qquad (4)$$

where $Q$ is the radiated energy per area of the disk surface, $G$ is the gravitational constant, and $M_{bh}$ is the black hole mass. Assuming that the disk material follows Keplerian orbits, i.e.

$$\omega = \left(\frac{GM_{bh}}{r^3}\right)^{1/2} \qquad (5)$$

from equations (3)-(4) one finds

$$Q(r) = \frac{3\dot{M}_{accr}GM_{bh}}{8\pi r_i^3}\left(\frac{r_i}{r}\right)^3\left[1 - \left(\frac{r_i}{r}\right)^{1/2}\right] \qquad (6)$$

The function $Q(r)$ [eq. 6] was originally derived by Shakura et al. [48] for binary systems with an accretion disk about a black hole. Further, assuming that the disk is emitting locally as a blackbody, the radial temperature distribution of the disk will be given by

$$T(r) = \left\{\frac{3\dot{M}_{accr}GM_{bh}}{8\pi r_i^3 \sigma}\right\}^{1/4}\left\{\left(\frac{r_i}{r}\right)^3\left[1 - \left(\frac{r_i}{r}\right)^{1/2}\right]\right\}^{1/4} \qquad (7)$$

where $\sigma$ is the Stefan-Boltzmann constant.

Defining the characteristic disk temperature $T^*$

$$T^* \equiv \left\{\frac{3\dot{M}_{accr}GM_{bh}}{8\pi r_i^3 \sigma}\right\}^{1/4} \qquad (8)$$

the expression of the radial temperature distribution becomes

$$T(r) = T^* \left\{ \left(\frac{r_i}{r}\right)^3 \left[1 - \left(\frac{r_i}{r}\right)^{1/2}\right] \right\}^{1/4} \qquad (9)$$

To further simplify mathematical expressions, we define the function $t(r/r_i)$

$$t(r/r_i) \equiv \left\{ \left(\frac{r_i}{r}\right)^3 \left[1 - \left(\frac{r_i}{r}\right)^{1/2}\right] \right\}^{1/4} \qquad (10)$$

Thus

$$T(r) = T^* t(r/r_i) \qquad (11)$$

Since $r_i$ is the inner disk radius, one has that $r \geq r_i$. It follows that

$$\max(t(r/r_i)) = t(7^2/6^2) = \frac{6^{3/2}}{7^{7/4}} \approx 0.488 \qquad (12)$$

Therefore the maximum temperature of a standard disk $T_{max}$ is given by

$$T_{max} \equiv \frac{6^{3/2}}{7^{7/4}} T^* \approx 0.488 \, T^* \qquad (13)$$

that is, the maximum surface temperature of a standard disk $T_{max}$ is approximately a half of the characteristic temperature $T^*$.

The total disk luminosity $L_{disk}$ is given by

$$L_{disk} = \int_{r_i}^{r_f} 4\pi r Q(r) dr \qquad (14)$$

where $r_f$ is the outer disk radius. Thus, substituting equation (6), one finds

$$L_{disk} = \frac{\dot{M}_{accr} G M_{bh}}{2 r_i} \left\{ 1 - \frac{3 r_i}{r_f} \left[ 1 - \frac{2}{3}\left(\frac{r_i}{r_f}\right)^{1/2} \right] \right\} \qquad (15)$$

The radial emission distribution of a standard disk [eqs. (6), (8), (10), (11)] are dependent on only four physical parameters, namely: black hole mass $M_{bh}$, mass accretion rate $\dot{M}_{accr}$, the innermost radius of the disk $r_i$, and the outermost radius of the disk $r_f$. The five assumptions used to derive the expressions for the radial emission distribution of a standard disk are: 1) conservation of angular momentum; 2) conservation of energy; 3) zero sheer stress at the innermost disk radius $r_i$; and 4) Keplerian orbits; and 5) local blackbody emission.

## 2.1 Additional Assumptions

We will apply two additional assumptions that reduce the dependence of the disk luminosity to only one parameter, namely: mass accretion rate $\dot{M}_{accr}$ .

. First, we shall assume that

$$r_f \gg r_i \qquad (16)$$

This assumption not only eliminates a free parameter, but it is justified since most of the radiation emission is coming from the inner disk regions [eq. (9)].That is, beyond a certain radius the disk contribution to the radiation continuum is negligible, even if the disk were to actually extend to infinity. The expression for disk luminosity [eq. (15)] becomes

$$L_{disk} = \frac{\dot{M}_{accr} GM_{bh}}{2r_i} \qquad (17)$$

Second, we shall also assume in this work, as was done by Shakura et al. [48] in their original accretion disk paper, and as is standard in the literature, that the innermost disk radius is given by the last stable circular orbit as determined by General Relativity (GR) under a Schwarzschild metric. That is

$$r_i = \frac{6GM_{bh}}{c^2} \qquad (18)$$

The expression for disk luminosity [eq. (17)], now becomes

$$L_{disk} = \frac{1}{12}\dot{M}_{accr}\, c^2 \qquad (19)$$

that is, assuming that the outer radius extends to infinity, and assuming the standard expression for the inner disk radius [eq. (18)], the efficiency of a standard disk is 1/12 . Note that, once again, the expression of $L_{disk}$ now depends on only one physical parameter, namely mass accretion rate $\dot{M}_{accr}$ . We note that since a real disk will have a finite outer radius, the exact value of the disk luminosity will depend on the exact position of the outer radius; however, if the disk is of sufficient size, then, in the calculation of the disk luminosity, the difference between taking into account the actual outer disk radius with respect to assuming an infinite disk would be negligible. Also, deviations of the inner disk radius from the standard expression [eq. (18)] may also generate corrections to equation (19).

## 2.2 Half-Thickness of the Disk $z_0$

The half-thickness of the accretion disk is determined by the point where the vertical component of gravity balances the thermal gas pressure and the continuum radiation pressure. That is, considering the corresponding hydrostatic equation along the vertical direction:

$$0 \approx -\rho \frac{GM_{bh}}{r^2} \frac{z_0}{r} - \frac{\partial P}{\partial z} + \rho \frac{\sigma_e Q}{c} \qquad (20)$$

where $\rho$ is the mass density of the disk at radius $r$, $P$ is the pressure, $\sigma_e$ is the electron scattering cross section per mass, $Q$ is the radiation flux coming from the disk surface, and $c$ is the speed of light.

### 2.2.1 $z_0$ for a Thermal-Pressure Dominated Disk

Assuming a region of the accretion disk where thermal-pressure dominates over continuum radiation pressure (i.e., continuum radiation pressure is negligible compared to thermal pressure), we have from equation (20):

$$0 \approx -\rho \frac{GM_{bh}}{r^2} \frac{z_0}{r} - \frac{\partial P}{\partial z} \qquad (21)$$

that we further approximate as:

$$\rho \frac{GM_{bh}}{r^2} \frac{z_0}{r} \approx \frac{P}{z_0} \qquad (22)$$

that is

$$\rho \frac{GM_{bh}}{r} \frac{z_0}{r^2} \approx \frac{P}{z_0} \qquad (23)$$

Since we are assuming here that the material in the disk is following Keplerian orbits (see §2), we have that

$$\frac{GM_{bh}}{r^2} = \frac{v_\phi^2}{r} \qquad (24)$$

Where $v_\phi$ is the Keplerian rotation speed at the radius r, therefore:

$$\frac{GM_{bh}}{r} = v_\phi^2 \qquad (25)$$

substituting in equation (23)

$$\rho v_\phi^2 \frac{z_0}{r^2} \approx \frac{P}{z_0} \qquad (26)$$

thus

$$v_\phi^2 \frac{z_o^2}{r^2} \approx \frac{P}{\rho} \qquad (27)$$

where $P/\rho$ is equal to the squared of the isothermal speed $v_{th,}$ therefore

$$v_\phi^2 \frac{z_o^2}{r^2} \approx v_{th}^2 \qquad (28)$$

we thus find that

$$z_0 \approx \frac{v_{th}}{v_\phi} r \qquad (29)$$

Now, assuming the ideal gas equation, and that the disk is predominately composed of ionized hydrogen, we have

$$P = 2nkT \qquad (30)$$

where $n$ is the number of hydrogen atoms per unit volume; therefore

$$P = 2\frac{\rho}{m_H} kT \qquad (31)$$

where $m_H$ is the mass of an atom of hydrogen; thus

$$\frac{P}{\rho} = \frac{2kT}{m_H} \qquad (32)$$

Therefore

$$v_{th} = \left(\frac{2kT}{m_H}\right)^{1/2} \qquad (33)$$

Thus

$$v_{th} = \left(\frac{2k}{m_H}\right)^{1/2} T^{1/2} \qquad (34)$$

substituting equation (9)

$$v_{th} = \left(\frac{2k}{m_H}\right)^{1/2} \left(T^* \left\{\left(\frac{r_i}{r}\right)^3 \left[1 - \left(\frac{r_i}{r}\right)^{1/2}\right]\right\}^{1/4}\right)^{1/2} \qquad (35)$$

that is

$$v_{th} = \left(\frac{2kT^*}{m_H}\right)^{1/2} \left(\frac{r_i}{r}\right)^{3/8} \left[1 - \left(\frac{r_i}{r}\right)^{1/2}\right]^{1/8} \quad (36)$$

Defining the characteristic isothermal sound speed $v_{th}^*$

$$v_{th}^* \equiv \left(\frac{2kT^*}{m_H}\right)^{1/2} \quad (37)$$

we have that:

$$v_{th} = v_{th}^* \left(\frac{r_i}{r}\right)^{3/8} \left[1 - \left(\frac{r_i}{r}\right)^{1/2}\right]^{1/8} \quad (38)$$

Since $r_i$ is the inner disk radius, one has that $r \geq r_i$. It follows that

$$\max(v_{th}(r/r_i)) = v_{th}(7^2/6^2) = \frac{6^{3/4}}{7^{7/8}} v_{th}^* \approx 0.699 \, v_{th}^* \quad (39)$$

Recalling that we assume here Keplerian orbits for mass inside the accretion disk, from equation (25)

$$v_\phi^2 = \frac{GM_{bh}}{r_i} \frac{r_i}{r} \quad (40)$$

thus

$$v_\phi = \left(\frac{GM_{bh}}{r_i}\right)^{1/2} \left(\frac{r_i}{r}\right)^{1/2} \quad (41)$$

Once again, since $r_i$ is the inner disk radius, one has that $r \geq r_i$. It follows that

$$v_{\phi max} = \left(\frac{GM_{bh}}{r_i}\right)^{1/2} \quad (42)$$

therefore

$$v_\phi = v_{\phi max} \left(\frac{r_i}{r}\right)^{1/2} \quad (43)$$

Substituting equations (38) and (43) in equation (29), we find

$$z_0 \approx \frac{v_{th}^* \left(\frac{r_i}{r}\right)^{3/8} \left[1 - \left(\frac{r_i}{r}\right)^{1/2}\right]^{1/8}}{v_{\phi max} \left(\frac{r_i}{r}\right)^{1/2}} r \quad (44)$$

thus

$$z_0 \approx \frac{v_{th}^* r_i}{v_{\phi max}} \left(\frac{r_i}{r}\right)^{-1/8} \left[1 - \left(\frac{r_i}{r}\right)^{1/2}\right]^{1/8} \left(\frac{r}{r_i}\right) \quad (45)$$

therefore

$$z_0 \approx \frac{v_{th}^* r_i}{v_{\phi max}} \left(\frac{r}{r_i}\right)^{9/8} \left[1 - \left(\frac{r_i}{r}\right)^{1/2}\right]^{1/8} \quad (46)$$

Defining the characteristic thermal-pressure disk-half-thickness $z_{0th}^*$

$$z_{0th}^* \equiv \frac{v_{th}^* r_i}{v_{\phi max}} \quad (47)$$

we obtain

$$z_0 \approx z_{0th}^* \left(\frac{r}{r_i}\right)^{9/8} \left[1 - \left(\frac{r_i}{r}\right)^{1/2}\right]^{1/8} \quad (48)$$

Analyzing the last equation, one finds that a thermal-pressure sustained accretion disk has a half-thickness that monotonously increases with radius.

### 2.2.2 $z_0$ for a Continuum-Radiation-Pressure Dominated Disk

Assuming a region of the accretion disk where continuum radiation dominates over thermal-pressure pressure (i.e., thermal pressure is negligible compared to continuum radiation pressure), we have from equation (20):

$$0 \approx -\rho \frac{GM_{bh}}{r^2} \frac{z_0}{r} + \rho \frac{\sigma_e Q}{c} \quad (49)$$

therefore

$$\frac{GM_{bh}}{r^2} \frac{z_0}{r} \approx \frac{\sigma_e Q}{c} \quad (50)$$

thus

$$z_0 \approx \frac{\sigma_e}{GM_{bh}c} Q r^3 \quad (51)$$

Substituting equation (6)

$$z_0 \approx \frac{\sigma_e}{GM_{bh}c} \frac{3\dot{M}_{accr}GM_{bh}}{8\pi r_i^3} \left(\frac{r_i}{r}\right)^3 \left[1 - \left(\frac{r_i}{r}\right)^{1/2}\right] r^3 \quad (52)$$

therefore

$$z_0 \approx \frac{3\sigma_e \dot{M}_{accr}}{8\pi c} \left[1 - \left(\frac{r_i}{r}\right)^{1/2}\right] \quad (53)$$

Defining the characteristic continuous-radiation-pressure disk-half-thickness $z_{or}^*$

$$z_{0r}^* \equiv \frac{3\sigma_e \dot{M}_{accr}}{8\pi c} \quad (54)$$

we obtain

$$z_0 \approx z_{0r}^* \left[1 - \left(\frac{r_i}{r}\right)^{1/2}\right] \quad (55)$$

From the last equation, one finds that a continuous-radiation supported disk, the half-thickness goes asymptotically to $z_{or}^*$ as the radius $r$ tends to infinity. In other words, except for the innermost regions of the accretion disk, a radiation supported disk can be approximated as a "flat" disk.

## 3. Accretion Disks within Typical QSO Parameters

Taking typical QSO parameters of black hole mass $M_{bh} = 10^9$ $M_\odot$ and disk Luminosity $L_{disk} = 10^{47}$ erg s$^{-1}$ (e.g., Hillier et al. [22]; Pereyra [38]), using equation (19), we find that the mass accretion rate for the disk $\dot{M}_{accr}$ is

$$\dot{M}_{accr} \approx 21.2 \; M_\odot \; \text{yr}^{-1} \quad (56)$$

following equation (18), the inner disk radius $r_i$ is

$$r_i \approx 8.86 \times 10^{14} \text{ cm} \approx 12{,}700 \; R_\odot \approx 59.2 \text{ AU} \quad (57)$$

Considering equation (8), the characteristic temperature of the disk $T^*$ is

$$T^* \approx 152{,}000 \text{ K} \qquad (58)$$

Thus, considering equation (13) the maximum temperature of the disk $T_{max}$ is

$$T_{max} \approx 74{,}300 \text{ K} \qquad (59)$$

For the disk temperature maximum obtained above, it implies the most of radiation coming directly from the disk is in the UV/Optical part of spectrum. In turn, reasonable fits have been found with the UV/Optical continuum of QSOs under the assumption of it being emitted directly from an accretion disk (e.g., Malkan [30]; Czerny et al. [11]; Wandel et al. [52]; Sun et al. [49]; Laor [26]; Krolik et al. [25]; Natali et al. [34]; Wilhite et al. [53]; Pereyra et al. [39]).

Additionally from equation (17), we note that half of the potential gravitational energy of the mass of the accretion disk is converted into emitted disk radiation, with the other half of this energy falling closer towards the black hole than the inner disk radius $r_i$. This energy that falls within the inner disk radius is likely converted through some physical process into X-ray emission. This last statement is supported by observations that show that the X-rays luminosities in QSOs tend to be comparable to the UV/Optical luminosities (e.g., Tananbaum et al. [50]; Grindlay et al. [19]; Laor et al. [27]; George et al. [15]).

The characteristic isothermal speed vth*, from equation (37), is

$$v_{th}^* \approx 50.1 \text{ km s}^{-1} \qquad (60)$$

Thus, considering equation (13) the maximum temperature of the disk $T_{max}$ is

$$v_{thmax} \approx 35.0 \text{ km s}^{-1} \qquad (61)$$

following equation (42)

$$v_{\phi max} \approx 122{,}000 \text{ km s}^{-1} \qquad (62)$$

Note that within the inner disk radius, we unavoidably enter a non-Newtonian regime; making QSOs, in principle, "potential labs" to test gravity theories beyond Newton's Gravitation.

The characteristic thermal-pressure disk-half-thickness $z_{0th}$, from equation (47), is

$$z_{0th}^* \approx 3.63 \times 10^{11} \text{ cm} \approx 5.21\, R_\odot \qquad (63)$$

Now, considering equation (54)

$$z_{0r}^* \approx 2.13 \times 10^{15} \text{ cm} \approx 30{,}600\, R_\odot \approx 142 \text{ AU} \qquad (64)$$

From the last two equations, we find that in the case of typical QSO parameters, the accretion disk is dominated by continuum radiation pressure. As implied from section §2.2.2, except for the innermost regions of the accretion disk, an accretion disk within a QSO can be approximated as a "flat" disk (see Fig. 1).

## 4. Estimates of Wind Mass Loss Rates from QSO Accretion Disk Winds

Pereyra et al. [37], showed through simplified one-dimensional models, that line-driven accretion disk winds were of steady nature. Pereyra et al. [37] analyzed several line-driven wind models, including one geared towards simulating disk wind models in general, one simulating inner disks winds, and one simulating outer disk winds (referred to as the "S", "I", and "O" models in Pereyra et al. [37]). Proving the steady of accretion disk winds was significant because, both cataclysmic variables (CVs) and QSOs show steady disk wind structures, and thus the steady nature of the steady nature of the wind becomes an important constraint for theoretical and computational models of these systems. In CVs, steady velocity structures with changes less than 10 km s$^{-1}$ persist for years (e.g., Fronning et al. [14]; Hartley et al. [20]). In QSOs, similar steady velocity structures persist for decades (e.g., Foltz et al. [13]; Barlow et al. [5]).

Additionally, Pereyra et al [37] found that for all three disk wind models they presented, that the wind mass loss rate $\dot{M}_{wind}$ was approximately equal to the "characteristic" wind mass loss rate $\dot{M}^*_{wind}$, that is

$$\dot{M}_{wind} \approx \dot{M}^*_{wind} \qquad (65)$$

In turn, the characteristic wind mass loss rate used in the models of Pereyra et al [37], was based on the mass loss rate of stellar line-driven winds (Castor et al. [10]), given by

$$\dot{M}^*_{wind} = \frac{B^* A^*}{\sigma_e v^*_{th}} \alpha (1-\alpha)^{\frac{1-\alpha}{\alpha}} k^{\frac{1}{\alpha}} \left(\frac{\gamma^*}{B^*}\right)^\alpha \qquad (66)$$

Where $B^*$ is the characteristic gravitational force per unit mass, $A^*$ is the characteristic area, $k$ and $\alpha$ are radiation line force parameters has initially defined by Castor et al. [10] in their study of line-driven stellar winds (see also Abbott [1]), and $\gamma^*$ is the characteristic radiation continuum force per unit mass.

As was seen in §3, the accretion disk in the case of QSOs is dominated by continuum radiation forces (rather than thermal pressure), and thus at surface the disk (see §2.2.2) the gravitational force from the supermassive black hole are at balance with the continuum radiation force; this in turn implies that

$$\gamma^* \approx B^* \qquad (67)$$

Substituting in equation (66), we find

$$\dot{M}^*_{wind} \approx \frac{B^* A^*}{\sigma_e v^*_{th}} \alpha (1-\alpha)^{\frac{1-\alpha}{\alpha}} k^{\frac{1}{\alpha}} \qquad (68)$$

The characteristic gravitational force per mass $B^*$, corresponds to gravitational force per mass near the disk surface along the direction of the wind streamline (that is, vertical), therefore

$$B^* \approx \frac{GM_{bh}}{r^2} \frac{z^*_{0r}}{r} \qquad (69)$$

Substituting in equation (68), we have

$$\dot{M}^*_{wind} \approx \frac{GM_{bh} z^*_{0r} A^*}{\sigma_e v^*_{th} r^3} \alpha(1-\alpha)^{\frac{1-\alpha}{\alpha}} k^{\frac{1}{\alpha}} \qquad (70)$$

That be rewritten in the form

$$\frac{\dot{M}^*_{wind}}{A^*} \approx \frac{GM_{bh} z^*_{0r}}{\sigma_e v^*_{th}} \alpha(1-\alpha)^{\frac{1-\alpha}{\alpha}} k^{\frac{1}{\alpha}} \frac{1}{r^3} \qquad (71)$$

In other words

$$\text{Wind mass loss rate per unit area} \approx \frac{GM_{bh} z^*_{0r}}{\sigma_e v^*_{th}} \alpha(1-\alpha)^{\frac{1-\alpha}{\alpha}} k^{\frac{1}{\alpha}} \frac{1}{r^3} \qquad (72)$$

We can thus estimate the total wind mass loss rate by integrating over the surface of the disk were the wind originates. Unlike the case of CVs, where the disk wind is likely coming from throughout the disk surface (Pereyra et al [40]); in QSOs the emission of X-rays coming from within the inner disk region (see §3) will ionize the material at the innermost surface disk areas such that the line-force there becomes negligible; as a consequence the disk winds in QSOs cannot start from the innermost regions of the disk, but from some point $r_{owind}$ ($> r_i$) and on, where the disk surface is shielded through some mechanism from the X-rays (e.g., Murray et al. [32]; Hillier et al. [22]; Pereyra [38]).

We thus have we can estimate the disk wind mass loss rate $\dot{M}_{wind}$ through the expression:

$$\dot{M}_{wind} \approx 2 \int_{r_{owind}}^{\infty} \frac{GM_{bh} z^*_{0r}}{\sigma_e v^*_{th}} \alpha(1-\alpha)^{\frac{1-\alpha}{\alpha}} k^{\frac{1}{\alpha}} \frac{1}{r^3} 2\pi r \, dr \qquad (73)$$

Note that the accretion disk has two surfaces (one "above" and one "below").

Therefore

$$\dot{M}_{wind} \approx 4\pi \frac{GM_{bh} z^*_{0r}}{\sigma_e v^*_{th}} \alpha(1-\alpha)^{\frac{1-\alpha}{\alpha}} k^{\frac{1}{\alpha}} \int_{r_{owind}}^{\infty} \frac{1}{r^2} dr \qquad (74)$$

Thus

$$\dot{M}_{wind} \approx 4\pi \frac{GM_{bh}z^*_{0r}}{\sigma_e v^*_{th}} \alpha(1-\alpha)^{\frac{1-\alpha}{\alpha}} k^{\frac{1}{\alpha}} \left[ -\frac{1}{r} \Big|^{\infty}_{r_{owind}} \right] \quad (75)$$

Therefore

$$\dot{M}_{wind} \approx 4\pi \frac{GM_{bh}z^*_{0r}}{\sigma_e v^*_{th}} \alpha(1-\alpha)^{\frac{1-\alpha}{\alpha}} k^{\frac{1}{\alpha}} \frac{1}{r_{owind}} \quad (76)$$

That is, the disk wind mass loss rate for a QSO can be estimated by the expression:

$$\dot{M}_{wind} \approx \frac{4\pi GM_{bh}z^*_{0r}}{\sigma_e v^*_{th} r_{owind}} \alpha(1-\alpha)^{\frac{1-\alpha}{\alpha}} k^{\frac{1}{\alpha}} \quad (77)$$

Using the values from §3 of black hole mass $M_{bh} = 10^9 M_\odot$ (same value used by Pereyra [38] in their 2.5D hydrodynamic simulations of QSO line-driven disk winds); the values of $z^*_{0r}$ and $v^*_{th}$ calculated in §3 (by assuming a disk luminosity of $L_{disk} = 10^{47}$ erg s$^{-1}$; value of disk luminosity also assumed in the models of Pereyra [38]); by taking the values used by Pereyra [38] for the line force parameters ($k = 0.002$ and $\alpha = 0.6$); by taking the value of $r_{owind} = 20r_i$ used by Pereyra [38]; and by applying equation (77) we find an estimate of the QSO disk wind mass loss rate of

$$\dot{M}_{wind} \approx 1.64 \times 10^{-2} M_\odot \text{yr}^{-1} \quad (78)$$

The above estimate value for $\dot{M}_{wind}$ is in very good agreement with the "exact" values of Hillier et al. [22] and Pereyra [38] of $\approx 10^{-2} M_\odot \text{yr}^{-1}$ (obtained through numerically intensive 2.5D time dependent hydrodynamic simulations)

## 4. Conclusions

As discussed in section §1 the accretion disk wind (ADW) scenario for QSOs continues to be a promising scenario to account for BAL QSOs as a consequence of viewing within a unified model for QSOs. Within the ADW scenario the outflows are coming from the accretion disk driven by line-radiation pressure from the disk. We show that in the ADW, for typical QSO parameters, the accretion is emitted mostly in the UV/Optical part of the spectrum, with the X-ray emission powered through some mechanism within the inner disk radius.

Thus within the ADW scenario, an X-ray shielding mechanism must be present in order not to over-ionize the gas above the disk, so that line-radiation pressure can be an effective wind driving force. That said, it is likely that the innermost regions of the disk are not X-ray shielded, and thus the disk would begin in a shielded region at a radius $r_{owind}$ greater than the inner disk radius. In this paper we also show that for typical QSO parameters, the accretion disk is dominated by continuum radiation pressure (rather than by thermal pressure), and that has a consequence of this the accretion disk in QSOs (except for the innermost regions) can be well approximated by a flat disk.

That said, the main result of this paper is the derivation of a relatively simple approximate expression for the QSO wind mass loss rates within the ADW scenario based on two fundamental QSO parameters: namely supermassive black hole mass and the disk mass accretion rate, and based on three fundamental QSO disk wind parameters: the standard CAK "$k$" and "$\alpha$" line-force parameters and the initial radius of the disk wind $r_{owind}$.

As we show above, the relative simple expression derived here for QSO disk wind mass loss rate is in very good agreement with the more "exact" values obtained through significantly more complex and detailed numerically intensive 2.5D time-dependent simulations.

**Figure Captions**

Fig. 1 Half-thickness of the accretion disk vs. radius for the typical QSO parameters used in this paper (black hole mass $M_{bh} = 10^9 \, M_\odot$ and disk Luminosity $L_{disk} = 10^{47}$ erg s$^{-1}$). Continuous line corresponds to the total disk half thickness ($z_{th} + z_r$). Dashed line corresponds to contribution to disk half-thickness ($z_r$) from continuum radiation pressure. Dotted line corresponds to contribution to disk half-thickness ($z_{th}$) from thermal pressure. As can been seen from the four plots (set at different scales on the *x*-axis), the continuum radiation pressure clearly dominates over thermal pressure up to several hundred times the inner disk radius $r_i$.(where almost of the disk radiation continuum is emitted). This in turn implies that the accretion disk in QSOs (expect for the innermost regions) can approximated as a "flat disk".

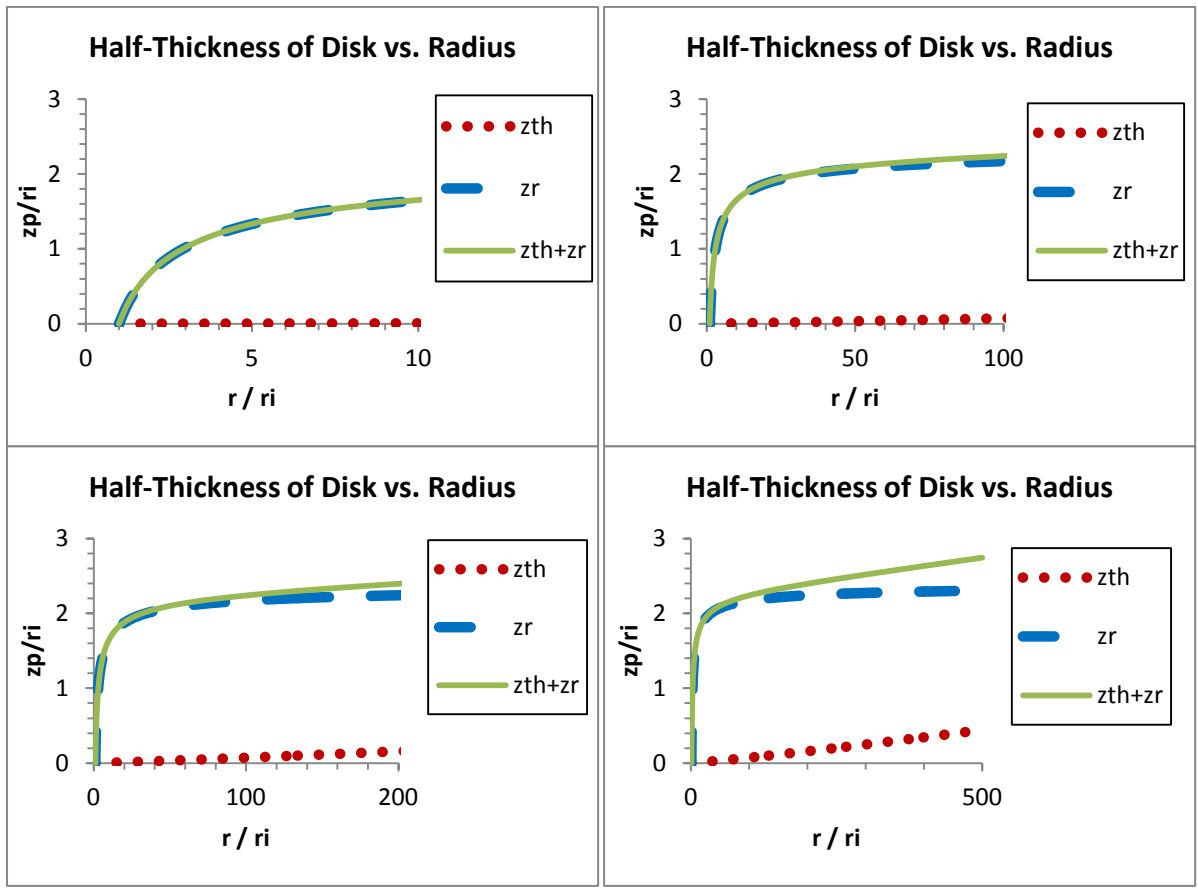

Fig. 1